\documentclass[aps,superscriptaddress,nofootinbib]{revtex4}
\usepackage{tikz,lipsum,lmodern}

\usepackage{amsmath}
\usepackage{slashed}
\usepackage{commath}
\usepackage{mathrsfs}
\usepackage{amsfonts}
\usepackage{amssymb}
\usepackage{graphicx} 
\usepackage{amsthm} 
\usepackage{float}

\newcommand{\lapproxeq}{\lower .7ex\hbox{$\;\stackrel{\textstyle<}{\sim}\;$}}
\newcommand{\gapproxeq}{\lower .7ex\hbox{$\;\stackrel{\textstyle>}{\sim}\;$}}
\newcommand{\stackdown}[2]{\lower 1.4ex\hbox{$\;\stackrel{\textstyle{#1}}
{\scriptstyle{#2}}\;$}}
\newcommand{\be}{\begin{equation}}
\newcommand{\ee}{\end{equation}}
\newcommand{\beq}{\begin{equation}}
\newcommand{\eeq}{\end{equation}}
\newcommand{\bea}{\begin{eqnarray}}
\newcommand{\eea}{\end{eqnarray}}

\makeatletter

\def\slash{\@ifnextchar[{\fmsl@sh}{\fmsl@sh[0mu]}}
\def\fmsl@sh[#1]#2{%
  \mathchoice
    {\@fmsl@sh\displaystyle{#1}{#2}}%
    {\@fmsl@sh\textstyle{#1}{#2}}%
    {\@fmsl@sh\scriptstyle{#1}{#2}}%
    {\@fmsl@sh\scriptscriptstyle{#1}{#2}}}
\def\@fmsl@sh#1#2#3{\m@th\ooalign{$\hfil#1\mkern#2/\hfil$\crcr$#1#3$}}
\makeatother

\def\beq{\begin{equation}}
\def\eeq{\end{equation}}

\catcode`\@=11 

\def\lsim{\mathrel{\mathpalette\@versim<}}
\def\gsim{\mathrel{\mathpalette\@versim>}}
\def\@versim#1#2{\vcenter{\offinterlineskip
    \ialign{$\m@th#1\hfil##\hfil$\crcr#2\crcr\sim\crcr } }}

\def\t1{{\tilde 1}}

\def\slash#1{#1\hskip-6pt/\hskip6pt}

\def\to{\rightarrow}

\begin{document}

\begin{flushright}
CERN-TH-2018-217\\
KCL-PH-TH/2018-45 \\
ACT-03-18, MI-TH-18-183
\end{flushright}
\vspace{0.2cm}

\title{Constraining D-foam via the 21-cm Line}

\author{John Ellis}

\affiliation{Theoretical Particle Physics and Cosmology Group, Department of Physics, King's College London,
Strand WC2R 2LS, London, U.K.}

\affiliation{National Institute of Chemical Physics and Biophysics, R\"avala 10, 10143 Tallinn, Estonia; \\
Theoretical Physics Department, CERN, CH-1211 Geneva 23, Switzerland.}

\author{Nick~E.~Mavromatos}

\affiliation{Theoretical Particle Physics and Cosmology Group, Department of Physics, King's College London,
Strand WC2R 2LS, London, U.K.}

\author{Dimitri V. Nanopoulos}

\affiliation{George P. and Cynthia W. Mitchell Institute for Fundamental
Physics and Astronomy, 
Texas A \& M University, College Station, TX 77843, USA}

\affiliation{Astroparticle Physics Group, Houston Advanced Research Center (HARC), \\ Mitchell Campus, Woodlands, TX 77381, USA.}

\affiliation{Division of Natural Sciences, Academy of Athens, Athens 106 79, Greece.}


\begin{abstract}

We have suggested earlier that D-particles, which are stringy space-time defects predicted in 
brane-inspired models of the Universe, might constitute a component
of dark matter, and that they might contribute to the masses of singlet fermions that could
provide another component. Interactions of the quantum-fluctuating D-particles with matter induce vector forces
that are mediated by a massless effective U(1) gauge field, the ``D-photon'', which is distinct from the ordinary photon
and has different properties from dark photons. We discuss the form of interactions of D-matter with conventional matter
induced by D-photon exchange and calculate their strength, which depends on the density of D-particles. Observations of the hydrogen
21~cm line at redshifts $\gtrsim 15$ can constrain these interactions and the density of D-matter in the early Universe.

\end{abstract}

\maketitle


\section{Introduction}

The nature of dark matter (DM) is one of the biggest mysteries in cosmology and particle physics.
It is commonly thought to be composed of one or more unknown species of particles, with
popular candidates ranging from ultralight bosons such as axions, through sterile neutrinos
and TeV-scale to supermassive metastable particles~\cite{BH}. Alternatively, the possibility that dark
matter might be composed of black holes has gained in interest following the recent
observations of gravitational waves emitted by mergers of black holes~\cite{LIGO}. Another current of
opinion is that dark matter might be due to some unexpected gravitational phenomenon~\cite{MOND}.
We have been pursuing a scenario for dark matter that is complementary to these approaches,
though with some aspects in common.

Our starting-point is an attempt to model quantum fluctuations in space-time - ``space-time foam"~\cite{wheeler} -
using elements derived from string theory~\cite{string}, in which the Universe may be modelled as a
three-brane world moving in a higher-dimensional bulk space~\cite{dfoam,westmuckett,sakharov,emnnewuncert,li,mavrolorentz}. Generic string models predict the appearance 
in this bulk space of point-like space-time defects called D-particles~\cite{polch}, such as D0-branes in Type IIA strings and
D3-branes wrapped around appropriate three-cycles in Type IIB strings. As the three-brane world and the 
D-particles move in the bulk, they may encounter each other, in which case an observer on the three-brane 
perceives the D-particle defects as flashing on and off, giving the 3+1-dimensional space-time a foamy structure.
In such a scenario, ordinary matter and radiation are represented by open strings with their ends attached on the three-brane.

Some of the D-particles may be trapped on the three-brane, in which case they would act as dark matter~\cite{vergou,sakel}.
However, depending on the dynamics of the bulk and the three-brane, the density of D-particles on the three-brane 
may evolve differently from the conventional dust-like dilution of matter density as the Universe expands.
 The D-particles interact with conventional matter particles via the capture and subsequent re-emission of
 open strings, accompanied by recoil of the D-particle~\cite{recoil}. Such interactions with D-particles could contribute 
 to the masses of singlet fermions that would also contribute to the dark matter density~\cite{emnsinglet}. The (quantum-fluctuating) D-foam 
 defects break Poincar{\' e} invariance, and their recoil during the interaction with propagating open-string states breaks 
 Lorentz invariance. The non-trivial momentum transfer during the interactions of matter strings with the 
 defects is mediated by non-local intermediate string states that do not admit a local effective action description, 
 leading to a violation of Lorentz invariance that is subject to probes using astrophysical sources~\cite{nature,emnnewuncert,li,review}.
 The recoil interaction of conventional matter with the D-particle defects can be described in terms of an 
 effective U(1) gauge field - the D-photon - whose strength depends on the density of D-particles~\cite{dfoam,westmuckett,emnnewuncert,li,mavrolorentz,emnsinglet}.
  
The purpose of this paper is to analyze in more detail the form of this D-particle/matter interaction, calculating its strength
and discussing constraints on its strength. We recall that there are important constraints 
on the density of dark matter and the strength of its interactions with conventional matter at the current epoch.
However, there are relatively few constraints from earlier cosmological epochs. The formation of 
astrophysical structures requires the presence of dark matter at redshifts $0 < z \lesssim 1000$.
The pattern of anisotropies in the cosmic microwave background radiation (CMB) is sensitive to the 
density of dark matter at a redshift $z \sim 1000$, and there are weak constraints on dark matter/ordinary 
matter interactions at this and higher redshifts. 

A new window on the interactions between dark matter and conventional matter has been opened
up by the possibility of measuring features in the spectrum of the 21-cm radiation emitted during
transitions between the triplet and singlet states of the hydrogen atom that occurred at redshifts $z \in (15, 20)$,
prior to the so-called cosmic dawn. The amplitude of such a prospective signal is given by
\begin{equation}
T_{21} = 35~{\rm mK} \, \Big(1 - \frac{T_\gamma}{T_s}\Big)\, \sqrt{\frac{1 + z}{18}}~,
\label{21signal}
\end{equation}
where $T_\gamma$ is the CMB temperature and $T_s$ is the singlet/triplet spin temperature of the hydrogen gas,
which is defined by the relative populations of the  lowest-energy spin-0 and spin-1 states. 
Standard arguments within the $\Lambda$CDM cosmological model imply the relation 
$T_\gamma \gg T_s \gsim T_{\rm gas}$ at $z \simeq 20$, where $T_{\rm gas}$ is the temperature of the 
hydrogen gas (essentially the baryon kinetic temperature, $T_k$), which is expected to be 
$T_{\rm gas} |_{z\simeq 17} \simeq 6.8 ~{\rm K}$, leading on the basis of (\ref{21signal})
to the expectation that $T_{21} > -0.2$. 

Interest in this new window on dark matter has been stimulated 
recently~\cite{barkana,recent1,recent,zurek,cooleqs,jia,houston,sikivie} by the claim of a strong absorption
feature  at redshifts $z \simeq 17$ in the sky-averaged 21-cm spectrum by the EDGES Collaboration~\cite{edges},
which is considerably larger then the standard $\Lambda$CDM expectation. However, we do not commit
ourselves to this interpretation of the EDGES data, which has not been accepted unanimously~\cite{Hills}. Rather, we discuss how this
Type of observation can be used to constrain the D-particle model for dark matter that we introduced above.

The constraint arises from the possibility of D-photon exchange between particles of dark matter and ordinary matter,
which would have effects similar to the exchange of a conventional photon or a (near-)massless dark photon~\cite{darkphoton}.
These would generate extra interactions of baryons with the dark matter fluid, beyond those predicted 
within the standard $\Lambda$CDM model. Such interactions would cause the baryons to lose kinetic energy 
and cool down more than in the $\Lambda$CDM framework. 

As a prototypical example of such a scenario, it was proposed in~\cite{barkana} that if the dark matter consists of 
more than a single dominant species, a fraction $f$ due to a millicharged component with mass $m_\chi$:
\begin{equation}\label{boundsf}
 (\frac{m_\chi}{{\rm MeV}}) \, 0.0115 \% \lesssim f \lesssim 0.4 ~\%
 \end{equation}
and an electric charge in the range $10^{-6}- 10^{-4}$ of the electron charge could be consistent with the 
baryon cooling interpretation of the EDGES data~\cite{cooleqs,zurek}.
 This could be consistent with other existing stringent constraints, including those stemming from big bang nucleosynthesis, 
 if $m_\chi$ is in a mass range much lower than the standard WIMP,  $m_\chi \in (5 - 35)$~MeV~\cite{recent1,recent}. 
However, some degree of fine-tuning would be needed in order to mitigate the effects of several astrophysical sources of 
baryon heating, which make it difficult to attain the level of baryon cooling required to explain the 21 cm signal~\cite{recent}.
Interactions of baryons with (millicharged) DM particles mediated by dark photons
that mix with ordinary photons~\cite{darkphoton} have also been considered in this context~\cite{jia}~\footnote{The stringent 
constraints on millicharged dark matter are avoided in models for baryon cooling via interactions with axion condensates~\cite{houston,sikivie},
either of Quantum Chromodynamics (QCD) origin~\cite{sikivie2}, or string-inspired~\cite{axiverse}, 
but these are subject to other astrophysical constraints.}. 

The framework we consider here is a multi-component dark matter model, in which
one of the components is a warm sterile Dirac neutrino with a mass in the range of a few tens of keV~\cite{ruffini}
with strong (compared to the weak Standard Model sector) self-interactions mediated by dark photons,whose r\^ole is played here by D-photons.
This contribution to the dark matter can play a crucial r\^ole in providing the observed 
halo-core structure of galaxies, and contribute towards the alleviation of discrepancies between predictions based on 
numerical simulations based on the $\Lambda$CDM model and observations at galactic scales 
(the so-called ``small-scale'' cosmology crisis~\cite{smallscale}). 

We discuss here the contribution to elastic scattering with such sterile neutrinos mediated by the exchange
of the effective D-photon U(1) gauge field induced by the recoil of D-particles, and calculate its magnitude.
We discuss how this interaction could contribute to baryon cooling, and show how any upper limit on such an effect could be used
to constrain the properties of D-foam.

The structure of the paper is as follows. In Section \ref{sec:dfoam} we review the main features of the stringy D-foam model, 
which offers a microscopic multi-species dark matter model framework within brane/string theory. 
In the following Section \ref{sec:21cm} we discuss the strength of the D-photon-mediated interactions between baryonic
matter and singlet sterile Dirac neutrinos, and how these interactions could contribute to baryon cooling. 
Our conclusions and outlook are summarised in Section~\ref{sec:concl}. In an Appendix we discuss 
some complimentary aspects of D-matter interactions with Standard Model particles, namely universal lensing properties of D-foam
originating from the back reaction on space-time of the recoiling D-particles during their interaction with generic matter.

\section{A Model for D-Foam Interactions with Singlet Fermions \label{sec:dfoam}}

\begin{figure}[ht]
\centering
\includegraphics[width=6.5cm]{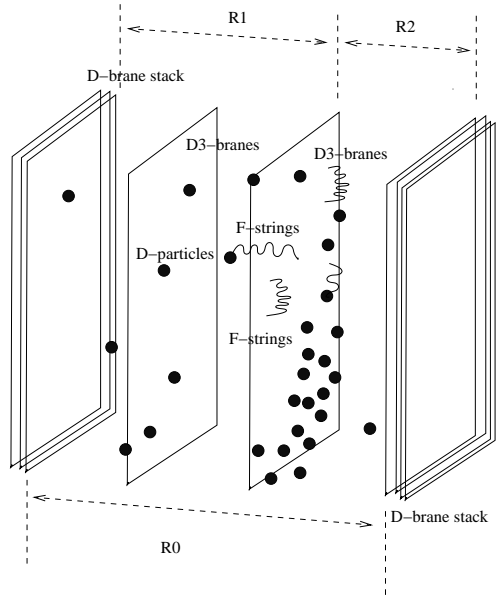}
\caption{\it Schematic
representation of a string/brane-theory-inspired prototype of a D-particle model of space-time foam~\cite{dfoam} proposed in~\cite{westmuckett}.
D-particles are represented by dots, and conventional particles are excitations on D3-branes that may be connected to D-particles.}%
\label{fig:dfoam}%
\end{figure}

We advocate a (toy) model of space-time foam~\cite{wheeler}, motivated by string 
theory~\cite{dfoam,westmuckett,sakharov,emnnewuncert,li,mavrolorentz}, in which
(effectively point-like) compactified brany stringy defects~\cite{string}, termed D-particles, provide foamy space-time
structures as illustrated in Fig.~\ref{fig:dfoam}.
Our Universe is described as a three-brane world moving in a higher-dimensional bulk space that is
punctured by D-particle defects. As the brane world and the D-particles move in the bulk, 
they may encounter each other, and an observer on the D3-brane would see the D-particle defects as flashing 
on and off, thus giving the space-time a foamy nature~\cite{wheeler}. 

In addition, some of the D-particles may be trapped on our three-brane, 
contributing to the density of cold dark matter. However,
the density of these defects is not necessarily uniform, and the 
effective density of D-particles on our three-brane world varies with the cosmic redshift in a model-dependent way
that is linked to the density profiles of the D-particles in the bulk space. For this reason, the D-particle density may not evolve
with redshift in the same way as conventional dust-like dark matter.

Depending on the string theory considered, the D-particles may either be point-like 
D0-branes (as in Type IIA strings)~\cite{dfoam,westmuckett,emnnewuncert} or D3-branes wrapped around 
appropriate three-cycles (as in Type IIB strings~\cite{li}). 
In such scenarios, ordinary matter and radiation are represented by open strings with their ends 
attached on the D3-brane.
The D-particles interact with Standard Model matter particles in a topologically non-trivial manner,
via the capture and re-emission of the open strings, accompanied by recoil of the D-particle,
as illustrated in Fig.~\ref{fig:dfoam}. The D-foam defects break Poincar\'e 
invariance, and these interactions - which are mediated by a non-local intermediate U(1) vector field -
cause Lorentz invariance to be broken in the propagation
of open string states.

In our approach~\cite{dfoam,emnnewuncert,li} the D-foam defects are to be thought of as providing an
environment or background medium, analogous to the lattice of ions in a solid-state system, and the
U(1) vector field mediating the interaction with matter particles is analogous to a phonon excitation~\cite{emnsinglet}.
Its vector nature is due to its r\^ole as a Goldstone boson associated with the spontaneous breaking of
Lorentz symmetry due to the recoiling D-particle defect.
We recall that the strength of the D-foam-matter interactions is not universal for different particle species.
In particular, the interactions are non-trivial only for photons and any other neutral particles that have 
no non-trivial internal quantum numbers~\cite{sakharov,emnnewuncert}, for which there is no obstacle to 
absorption and re-emission by a D-particle~\footnote{A form of D-brane dark matter has also been proposed
independently in \cite{shiu}, but from a rather different perspective. In particular, that model did not have 
non-universal couplings to different matter species.}.

It is well known that, in string theory, electrodynamics is described by a Born-Infeld extension of the
minimal Maxwell Lagrangian of conventional QED with photon field $A_\mu$ and field strength $F_{\mu \nu}$. Moreover,
it was argued in \cite{vergou,sakel,emnsinglet} that the effective low-energy theory describing
the interaction of the D-foam with matter strings is also represented by Born-Infeld-Type dynamics for the D-photon
field ${\mathcal A}_\mu$. Their combination,
including the D-photon interaction with a fermion current, can be written in a manifestly dual form~\cite{bi2field} as:
\begin{eqnarray} \label{bi1}
{\mathcal L}_{BI}^{\rm two~fields} & = & \frac{T^2}{g_s} \sqrt{ 1  + \frac{T^{-2}}{4} (e^2\, F_{\mu\nu}F^{\mu\nu}  + \frac{g_s}{4} {\mathcal F}_{\mu\nu} {\mathcal F}^{\mu\nu}) 
+ \frac{T^{-4}}{16} \Big( e \, g_s^{1/2}\, F_{\mu\nu} \, \epsilon^{\mu\nu\rho\sigma} {\mathcal F}_{\rho\sigma} \Big)^2 + \dots } - \frac{T^2}{g_s}
 \nonumber \\
&+& \sum_{I=b,N} \, \int d^4 x \, {\tilde g_{V,I}} \, {\mathcal A}_\mu \overline \psi_I \gamma^\mu \, \psi_I + \sum_{I=b,N}\, \int d^4 x \, \overline \psi_I i \, \slash{\partial} \psi_I   + \dots 
\end{eqnarray}
in the notation of \cite{emnsinglet}, where we refer the reader for details. Here 
the quantity  ${\mathcal F}_{\mu\nu}=\partial_\mu {\mathcal A}_\nu - \partial_\nu {\mathcal A}_\mu$
denotes the field strength tensor of the  vector field ${\mathcal A}_\mu$,
$T \equiv 1/(2\pi \alpha^\prime )$ where $\alpha^\prime = M_s^{-2}$ is the string tension with
$M_s$ the string mass scale, which is in general lower than the four-dimensional Planck mass~\cite{string},
$g_s$ is the string coupling~\footnote{It should be understood that in the presence of a non-trivial, 
space-time coordinate-dependent dilaton field, $\Phi (x)$, 
the string coupling $g_s$  in (\ref{bi1}) would be replaced by $e^{\Phi (x)} g_s$.}, the $\dots$ in the upper line denote $B$-field terms,
and the $\dots$ in the lower line denote other fields as well as the higher-derivative terms that appear in string effective actions.
The fermion fields $\psi_I$, $I=b, N$ may represent either a baryon ($I=b$) or a singlet fermion such as a sterile neutrino ($I=N$).
The reader should remember that in open string theory~\cite{string}, the electron charge $e$ is expressed  in terms of the string coupling  as:
\begin{equation}\label{gse}
g_s = e^2 \, ,
\end{equation}
which stems from the requirement that the electromagnetic part of the low-energy Born-Infeld terms in the target-space-time Lagrangian 
that describes the propagation of open strings 
in an electromagnetic background should yield the canonically-normalised Maxwell action to lowest order in derivatives~\footnote{We
note that any mixing between the D-photon and the ordinary photon in (\ref{bi1}) occurs at fourth order in the derivative expansion. Thus, 
there is no kinetic mixing at second-derivative order, as in standard dark photon models. Truncating the derivative expansion of the 
Born-Infeld square root to second order is sufficient for our purposes, in which case D-photon/photon mixing can be neglected.}. 

As explained in \cite{mavrolorentz,emnsinglet}, in addition to the recoil vector contributions in (\ref{bi1}), there are also contributions arising from 
the antisymmetric (spin-one) tensor field $\widehat B_{\mu\nu}$ in the massless string multiplet.
In our approach (through summation of world-sheet genera to encapsulate the quantum string loop effects properly),
this has been argued to be a derivative operator whose components are given by
\bea\label{btensor}
\widehat{B}_{i0} = \widehat{u}_i = -i g_s \, \frac{r_i}{M_s} \,\frac{\partial}{\partial X^i} &\equiv & -i g_s \, \frac{r_i}{M_s}\, \nabla_i
\;  ({\rm no~sum~over~} i = 1,2,3) ~, \nonumber \\
 \widehat{B}_{ij} = \epsilon_{ijk}\,\widehat{u}^k~,
\eea
where $r_i$ denotes the fraction of momentum transferred during each collision of a matter string with a D-particle defect.   
Making a derivative expansion of the effective action, 
and averaging stochastically  over the population of D-particles, with $\langle r_i \rangle = 0$, and $\langle r_i r_j \rangle = \sigma^2 \, \delta_{ij}$, 
these terms 
yield quartic (and higher-order) Lorentz-violating terms in the Lagrangian of the form 
\begin{equation}\label{nablaint}
{\mathcal L}_{\rm B-field} \, \ni  \, \frac{1}{16} {\mathcal F}_{\mu\nu}\, 
\frac{g_s^2 \tilde \sigma^2}{\,M_s^2} \Delta  \, {\mathcal F}^{\mu\nu}, 
\end{equation}
where $\Delta$ is the 3-space Laplacian, $\Delta \equiv \nabla_i \nabla^i = \vec{\nabla} \cdot \vec{\nabla}$. 
The quantity $\tilde \sigma^2 \propto \sigma^2$, where the numerical proportionality factor is associated with 
quantum ordering ambiguities~\cite{mavrolorentz}, and can be absorbed in the definition of the stochastic variance.
This is treated here as a phenomenological parameter characterising the foam, which depends in general on its density.
It was argued in \cite{emnsinglet} that these terms would generate dynamical masses for the singlet fermions, 
as discussed briefly below.

 The baryons $\psi_b$ are Dirac fermions. On the other hand, depending on the underlying microscopic model, 
 the singlet fermions $\psi_N$ could be either Dirac or Majorana. In the latter case, the Majorana 
fermion field $\psi_N^M$ could have only axial-vector couplings with the ${\mathcal A}$ vector field,
as follows from charge conjugation of the the Majorana field. Therefore, the effect of the dark matter particles on
baryon cooling through D-photon exchanges would be negligible in the Majorana case~\footnote{Moreover, the
cooling of baryons through their direct interaction with the D-foam would in general be too weak
to be of phenomenological interest, though condensates of D-particles could enhance the effect,
as discussed briefly at the end of Section \ref{sec:21cm}.}, unless Standard Model fermions ($f$) 
also have axial couplings $\tilde g_{A,I}$, $I=f,N$ with the D-photon ${\mathcal A}_\mu$, mimicking $Z$-portal
models for DM~\cite{zportal}:
\begin{equation}
{\mathcal L}_{\rm fermion-D-foam-axial} \ni \sum_{I=b,N} \, \int d^4 x \, {\tilde g_{A,I}} \, {\mathcal A}_\mu \overline \psi_I \gamma^\mu \, \gamma^5 \, \psi_I ~.
\label{axial}
\end{equation}
The analysis of elastic scattering involving axial current couplings to DM and nucleons in the Majorana 
case is entirely analogous to the case of Dirac DM, which we consider below for concreteness.

Thus, in the case of either Dirac or Majorana singlet DM fermions, 
vector or axial vector interactions between the ${\mathcal A}$-field 
and the fermions are allowed, as in (\ref{bi1}), (\ref{axial}), and this may lead to an interesting mechanism for baryon cooling through the
elastic scattering of baryons with singlet fermions, via tree-level ${\mathcal A}$ exchange, mimicking, straightforwardly, 
the millicharged DM case. But such a similarity 
is only formal, given that the $D$-photon coupling is not an electromagnetic one, as already explained. 

In what follows we shall restrict for concreteness our discussion to the Dirac DM fermion case. 
We note at this point, that, as follows from our formalism (\ref{bi1}), Dirac antifermions couple to the ${\mathcal A}$-field with 
couplings $\tilde g_{V (A), I}$ of opposite sign. This stems from the CPT properties of the Dirac equation
in the low-energy effective action (\ref{bi1}), and is consistent with the gauge nature of the  
recoil excitations of the D-particles, which are represented by open strings stretched between the D-particle defect and the brane in the T-dual formalism~\cite{recoil}. For Majorana particles, on the other hand, the axial coupling (\ref{axial}) guarantees that particles and antiparticles have identical couplings.
We note that, since the D-particle recoil interaction is mediated by a spin-one open string excitation, in general it need not be the same for
particles and antiparticles. This is indeed the case for a Dirac DM fermion, which may have either vector and/or axial-vector recoil interactions~\cite{li}.

The couplings~\footnote{In what follows we restrict ourselves for concreteness to vector couplings, however entirely analogous considerations characterise the axial D-photon coupling to fermions.} 
${\tilde g}_{V, I}$, $I=b,N$ are in general of different strengths, as discussed above, in particular ${\tilde g}_{V b} =0$ in Type IA models.  
The factor $g_s^{1/2}$ pulled out explicitly in front of ${\mathcal F}$-terms in the argument of the square root in (\ref{bi1}) 
ensures the canonical normalisation of the Maxwell-like kinetic term of the ${\mathcal A}_\mu$ gauge field, 
when the square root is expanded in powers of $T^{-1}$. Since we are interested here only in low-energy dynamics,
we neglect the higher-order terms in ${\mathcal F}_{\mu\nu}$ in (\ref{bi1}), and restrict our attention to the leading-order Maxwell term.

To summarise, the relevant low-energy part of the effective target-space action on our three-brane world, where the quantum fluctuating D-particle 
meets the open-string singlet fermion state, is described by the Lagrangian (\ref{bi1}) expanded in powers of derivatives: 
\begin{eqnarray}\label{bi}
S =  \int d^4 x   \left[\frac{1}{4} {\mathcal F}_{\mu\nu}\left(1
 + \frac{1}{4}\, \frac{g_s^2 \tilde \sigma^2}{\,M_s^2} \Delta \right) \, {\mathcal F}^{\mu\nu} \right] \, 
+ \sum_{I=b,N} \, \int d^4 x \, {\tilde g_{V,I}} \, {\mathcal A}_\mu \overline \psi_I \gamma^\mu \, \psi_I + \sum_{I=b,N}\, \int d^4 x \, \overline \psi_I i \, \slash{\partial} \psi_I
 + \, \dots ~,
 \end{eqnarray}
where  the $\dots$ represent terms of higher order in derivatives as well as elecromagnetic and other fields in the string effective action. 
The Lorentz-violating terms in (\ref{bi}) that are quartic in derivatives 
are relevant for the generation of non-perturbative masses for singlet fermions $\psi_I$, 
as discussed in~\cite{mavrolorentz,emnsinglet}:
\bea
\label{dynmass2}
m_{dyn}^I \simeq M {\rm exp}\left(-\frac{2\pi}{3\,\alpha_{V, I}}\right)~, \quad I=b, N, \quad 
 \alpha_{V, I} = \frac{{\tilde g}_{V,I}^2}{4\pi}~, \quad M = \frac{M_s}{\,g_s \,\sqrt{{\tilde \sigma}^2}}~,
 \eea
 but do not play any r\^ole in baryon cooling, 

The magnitude of the dimensionless coupling ${\tilde g}_V$ depends on the string model.
As discussed in \cite{emnsinglet}, in the context of the Type IIB model of D-foam~\cite{li}, in which
our world is viewed as a D7 brane with four dimensions compactified, 
and the ``D-particles'' of the foam are represented by compactified D3 branes wrapped around three cycles, we find for the 
coupling of the D-foam to singlet fermions
\bea\label{form}
\tilde g_{V,N} \propto g_s^{1/2} \sqrt{n_D^{(3)} \, {R^\prime}^{-1} } \, {\mathcal F}(s,t, \alpha^\prime) \, M_s^{-2}~,
\eea
where $n_D^{(3)}$ is the density of D-particles, and $R^\prime $
is the radius of the fourth space dimension of the D7 brane transverse to the D3 brane.
A phenomenological estimate is $R^\prime \sim 338 \, M_s^{-1}$ in the model of \cite{li}, but different values are possible in other models. The quantities 
$s$ and $t$ in (\ref{form}) are Mandelstam variables
and ${\mathcal F}(s,t, \alpha^\prime)$ is a momentum-dependent form factor associated with string 
amplitudes describing the scattering of such singlet fermionic excitations off D-particles in the model, 
including string-loop corrections that are suppressed by powers of $g_s$
that modify the coupling $\tilde g_V$.  This form factor is difficult to compute exactly,  given 
that the target-space action of D-branes is not fully known. However, for the slowly-moving excitations 
with momenta that are small compared to the string mass scale $M_s$ that are of relevance here,
a field-theoretical approximation is fully adequate, and the form factor is well approximated by unity.
  
In the Type IIB string-foam model of \cite{li}, electrically-charged baryons (or electrons)
interact with the recoiling D-foam with a coupling that is very suppressed to (\ref{form}):
 \bea\label{formb}
\tilde g_{V,b}  = (1.55 \, M_s^{-1})^{2}\, \sqrt{n_D^{(3)} \, {R^\prime}^{-1}} \, \tilde g_{V, N} 
\propto g_s^{1/2} \Big(n_D^{(3)} \, {R^\prime}^{-1} \Big)\, (1.55 \, M_s^{-1})^{2} \ll {\tilde g}_{V,N}~.
\eea  
We note in passing that in \cite{emnsinglet} we discussed a geometric mechanism that enhances the dynamical 
fermion mass (\ref{dynmass2}) via a suitable embedding of the model in a higher-dimensional set-up
involving brane worlds~\cite{MR} in a Randall-Sundrum (RS) warped bulk geometry~\cite{RS}.
By selecting appropriately the parameters of such an embedding one can generate a mass for the sterile neutrinos
that is of phenomenological interest~\footnote{However, Standard Model fermions acquire their
masses from the conventional Higgs mechanism.}. Thus, it was argued in~\cite{mavrolorentz,emnsinglet} that another observable implication of D-foam
could be the dynamical generation of small non-perturbative masses for singlet neutral fermions in Type IA  D-foam models,
where the foam is transparent to charged fermions, and for both singlet neutral and charged fermions in Type IIB D-foam models,
though the charged and singlet fermions couple with different strengths to the D-foam. 

We shall not discuss dark energy here, assuming that it is small at the epoch of interest. 
In the presence of a non-trivial, space-time-coordinate dependent dilaton field, $\Phi (x)$, 
the string coupling $g_s$  in (\ref{bi1}) would be replaced by $e^{\Phi (x)} g_s$.
However, we assume that the value of the dilaton field $\Phi$ is constant during this era, 
and in general at later epochs of the Universe, and normalise it to zero.

\section{Baryon Cooling in D-foam \label{sec:21cm}}

In this Section we consider the possible contribution of D-foam to baryon cooling,
focusing on the Type IIB model of \cite{li}. The effect in 
Type I string D-foam is strongly suppressed because there is no direct interaction of the D-photon with the baryon (or charged matter more generally).
In this model, the only way to generate an interaction between dark matter and ordinary matter would be through the kinetic mixing of the D-photon with
the ordinary photon. However, in our Born-Infeld model (\ref{bi1}) this happens only at four-derivative order, and
involves two D-photons and two ordinary photons. In the Type IIB model, a D-photon mediator can directly couple with a baryon,
yielding a $v^{-4}$ Rutherford-like behaviour of the dark matter/baryon scattering cross-section.
This can be significant if the dark matter has a warm component of sterile neutrinos, with mass a few tens of keV, as in the model of \cite{ruffini}.
In this case, the reduced mass $\mu_{N,b}$ will be of order of the sterile neutrino mass, which is  much smaller than the baryon mass, 
so the cross-section can be much larger than that for other possible dark matter species with 
masses $m_\chi \gg 1~{\rm GeV}$ that might interact with the baryons, thereby providing the leading cooling mechanism for baryons, beyond the standard cosmological model, leading to strong
potential constraints on the D-foam parameters as we discuss below~\footnote{As was already stressed in the previous Section,
the D-particles, with masses $M_s/g_s \sim M_s/(4\, \pi\, \alpha) \gg $~TeV are assumed not to propagate like ordinary excitations.
Instead, in a low-energy limit, their interaction with ordinary particle excitations is expressed via
the effective recoil vector field ${\mathcal A}_\mu$ that mediates dark matter/baryon interactions.
However, this D-foam background may exhibit non-trivial optical properties (refractive indices), which have been studied 
elsewhere~\cite{review,emnnewuncert,li} and provide complementary phenomenological constraints on the D-foam model.}. 

Using the Lagrangian (\ref{bi}) and ignoring higher-derivative Lorentz-violating terms,  
one can readily calculate the total cross-section for elastic scattering of sterile neutrinos on baryons, 
which is mediated by tree-level ${\mathcal A}_\mu$ exchange, and and has the same form as the QED
Rutherford cross-section. It may be written in the form~\cite{cooleqs,recent1}  
\begin{equation}\label{sigmat1}
\sigma_t = \int d({\rm cos}\theta)\, (1 - {\rm cos}\theta) \frac{d \sigma(\theta)}{d({\rm cos}  \theta)} \, ,
\end{equation}
where $\theta$ is the scattering angle, and 
\begin{equation}\label{dcs}
 \frac{d \sigma(\theta)}{d ({\rm cos}\theta )}= -\frac{1}{2\pi} \int_0^{2\pi} \frac{d \sigma}{d\Omega} \, d\phi \, , 
 \end{equation}
where $d\sigma/d\Omega$ is the differential cross-section in the solid angle $\Omega$,
which  is inversely proportional to the fourth power of $v$, exactly as in the case of millicharged dark matter/-baryon scattering~\cite{recent}, 
but with the electric charge of the baryon replaced by the coupling $\tilde g_{V,b}$ (\ref{formb}), and the charge 
$\epsilon e$ of the millicharged dark matter particle replaced by $\tilde g_{V,N} \gg  {\tilde g}_{V,b}$:
\begin{equation}\label{sigmat}
\sigma_t \simeq \frac{2\pi \alpha_{V,N}^2 \,  \Big(n_D^{(3)} \, {R^\prime}^{-1} \Big) \, (1.55 \, M_s^{-1})^{4} }{\mu_{N,b} ^2 \, v^4} \, \xi  = 
\frac{2\pi \, \alpha^2 \,  \Big(n_D^{(3)} \, {R^\prime}^{-1} \Big)^2 \, (1.55 \, M_s^{-2})^{4} }{\mu_{N,b} ^2 \, v^4} \, \xi ~, 
\end{equation}
where $\mu_{N,b} = \frac{m_{N} m_b}{m_N + m_b}$ is the reduced mass of the sterile neutrino/baryon system,
$m_N$ ($m_b$) being the mass of the sterile neutrino (baryon). The quantity
 $\alpha_{V,N}$ was defined in (\ref{dynmass2}), and $\alpha = \frac{e^2}{4\pi}$ is the fine structure constant of electromagnetism. 
 In expressing the result in terms of $\alpha$, we took into account 
(\ref{gse}). The quantity  $\xi$ in (\ref{sigmat}) corresponds to the 
logarithmic Debye regulator function for the forward divergence of the momentum transfer integral in QED~\cite{zurek}, which we discuss below. 

Comparing the cross-section (\ref{sigmat}) with the corresponding one for millicharged dark matter/baryon interactions~\cite{recent}, 
the role of the  millicharge parameter $\epsilon = e_\chi/e$ is played here by the quantity 
\begin{equation}\label{neweps}
\varepsilon \equiv \Big(n_D^{(3)} \, {R^\prime}^{-1} \Big)^{1/2} \, (1.55 \, M_s^{-1})~, 
\end{equation}
in terms of which the cross-section (\ref{sigmat}) acquires a form similar to that for millicharged dark matter/baryon scattering~\cite{recent}:
\begin{equation}\label{sigmat2}
\sigma_t \simeq 
\frac{2\pi \, \alpha^2 \,  \varepsilon^4 }{\mu_{N,b} ^2 \, v^4} \, \xi ~.
\end{equation}
However, in contrast to the millicharged case, here both the baryon and the singlet fermion 
carry ``charge'' $\varepsilon e $ for the D-photon U(1) gauge group. 
The quantity $\varepsilon$ depends on the density of the D-foam $n_D (z)$, which is in general redshift-dependent,
as mentioned earlier, and the compactification parameters of the underlying  Type IIB string  theory model~\cite{li}.
Constraints on baryon cooling at redshifts $z \simeq 15$ to 20~\cite{edges} can be used to bound these model parameters.

Before discussing this, we first discuss the Debye function $\xi$ appearing in (\ref{sigmat}). 
As in the QED case, the differential cross-section 
\begin{equation}\label{logdebye}
\frac{d \sigma}{d ({\rm cos}\theta)}   = 
\frac{2\pi \, \varepsilon^4 \, \alpha^2 }{\mu_{N,b} ^2 \, v^4} \, \Big(\frac{1}{1 - {\rm cos} \theta}\Big)^2
\end{equation}
diverges when the forward scattering angle $\theta \to 0$. In the QED case this regularized~\cite{zurek} by the
Debye screening due to free electrons and protons in the plasma at redshifts $z \lesssim 20$, 
with a density $n_e=n_p = \Omega_b \rho_c a_R^{-3}/m_p$, where $\Omega_b$ is the baryon density 
normalised with respect to the critical density $\rho_c$, $a_R$ is the Universe scale factor at recombination, and $m_p$ is the proton mass. 
In millicharged dark matter scenarios, it is assumed that the dark matter is tightly coupled to the photon-baryon plasma
by Coulomb scattering due to photon exchange, so that the dark matter density fluctuations are damped.

In our D-foam case, although although the dark matter particle is electrically neutral,
it has a D-foam U(1) ``charge'' $\varepsilon e$, while  
electrons and baryons carry both the electric $e$ and the D-foam U(1) charge $\varepsilon^2 e$, 
see (\ref{formb}), (\ref{neweps}). Since the
coupling of the dark matter to baryons provided by this charge is dependent on the density of foam, constraints on the latter can be derived by  
assuming that it is not strong enough to affect the relative velocity difference between DM and baryons prior to recombination
as in the millicharged case~\cite{recent1}, or damp significantly the Baryon Acoustic Oscillations~\cite{sikivie}, which would lead to inconsistencies with observations.

Under this assumption, one can adapt the regularisation procedure of \cite{zurek} to this case, 
The divergence for forward scattering encountered in (\ref{logdebye}) is removed by
replacing the zero of the scattering angle $\theta \rightarrow 0$ by a minimum non-zero angle $\theta_{\rm min}  \ll 1 $, 
which is related to a maximum impact parameter $b_{\rm max}$  for dark matter/baryon scattering~\cite{zurek}:
\begin{equation}\label{maxscat}
b_{\rm max} = \frac{\varepsilon^3 \, \alpha}{\langle \mu_{N,b} v^2\rangle}\, {\rm cot}(\frac{\theta_{\rm min}}{2}) \quad  \stackrel{\theta_{\rm min} \ll 1}{\simeq}   
\quad  \frac{\varepsilon^3 \, \alpha}{3T}\, \frac{2}{\theta_{\rm min}}~,
\end{equation} 
where we have used $\langle \mu_{N,b} v^2\rangle = 3T$, as appropriate for thermalised particles. In the case of
standard electromagnetic scattering, one would equate this maximum 
impact parameter to the corresponding Debye length $\lambda_D$ of the baryon plasma, 
which indicates the maximum range of the electrostatic effect of a charged carrier. In our case,
the analogue of the Debye length for the effective U(1) interaction is not known: the full Born-Infeld
Lagrangian (\ref{bi1}) replaces the Maxwell part of the action (\ref{bi}), together with higher-order covariant-derivative terms acting on fermions, 
whose full form depends on the specific microscopic string model considered and cannot be expressed in closed form.  
Nonetheless, one can make the simplifying assumption that the U(1)
interactions of the dark matter with the baryons would mimic the standard electromagnetic
expression for the Debye length for a baryon population (``plasma'') in thermal equilibrium with the D-foam at a temperature 
$T$~\footnote{This is a rather drastic assumption, but is motivated by the fact that thermalised D-matter reproduces the 
correct large-scale structure of the Universe, if the D-matter constitutes a dominant dark matter component at large scales~\cite{sakel}.}. 
This follows from the fact that a Debye length scale 
\begin{equation}\label{ld}
\lambda_D = \Big(\frac{\epsilon T}{\sum_{j=1}^N n_j q_j^2}\Big)^{1/2},
\end{equation}
arises in thermodynamic descriptions of large systems consisting of $N$ charged species with mobile charges
$q_j$ and (number) densities $n_j$ in a medium with static permittivity $\epsilon$, 
following the corresponding Poisson equation. The similarity of the low-energy theory of the effective
D-foam interactions with standard electromagnetism implies that in our D-foam case,
where the baryons have U(1) charge $\varepsilon^2 e$ ((\ref{formb}), (\ref{neweps})), and density $n_p(z)$ 
at redshift $z$, the corresponding Debye screening length for D-foam effects in a baryon plasma at temperature $T_b(z)$, would be 
\begin{equation}\label{ldd}
\lambda^{\rm D-foam}_D = \Big(\frac{\epsilon\, T_b(z)}{4\pi \, \varepsilon^4\, n_b(z) \,\alpha}\Big)^{1/2} \, .
\end{equation}
Identifying $b_{\rm max} = \lambda^{\rm D-foam}_D$, and setting the D-foam vacuum permittivity $\epsilon=1$,
we obtain from  (\ref{maxscat}) and (\ref{ldd}):
\begin{equation}\label{thmin}
\theta_{\rm min} = \frac{2\, \varepsilon^3 \, \alpha}{3T_b \, \lambda^{\rm D-foam}_D } = \frac{2\, \varepsilon^5 \, \alpha^{3/2} \, ( 4\pi \, n_b)^{1/2}}{3\, T_b^{3/2}} \, .
\end{equation}
Performing the regularised integration over $\theta$ in (\ref{sigmat1}), using (\ref{logdebye}) and (\ref{thmin}), we find the following
expression for the logarithmic Debye regulator function $\xi$
\begin{equation}\label{logdebye2}
\xi =  {\rm ln}\Big(\frac{4}{\theta_{\rm min}^2}\Big) = {\rm ln}\Big(\frac{9\, T_b^{3}}{4\pi \, \varepsilon^{10} \, \alpha^3 \, n_b }\Big) =  {\rm ln}\Big(\frac{9\, T_b^{3}}{4\pi \, \alpha^3\, n_b }\Big) - 10 \, {\rm ln}\,\varepsilon~
\end{equation}
in our case.

We note some crucial differences from the millicharged case. In the latter, the interaction between DM and baryons occurs through 
the conventional photon of electromagnetism, and this is why only highly-ionising matter is relevant in that case. 
This implies that only the free-electron plasma fraction $x_e n_H$, where $x_e = n_e/n_H$, of the total density of the hydrogen gas 
$n_H$  interacts with DM, and hence in such a case one should replace $n_b$ in (\ref{logdebye2}) by $x_e n_H$. 
This fraction becomes very small during the cosmic dawn era: $x_e^{\rm dawn} \sim 2 \times 10^{-4}$~\cite{recent}, 
which is why in the millicharged DM case one would need a cross-section that exceeds the CMB limits~\cite{planck} by 
several orders of magnitude in order to reproduce the EDGES signal, assuming that millicharged particles provide
all the DM. Otherwise, if the millicharged DM constitutes only a fraction $f_\chi < 0.4\%$ of the DM, there is a small window in the 
parameter space of millicharged DM that reproduces the reported  EDGES signal while being
consistent with the CMB constraint (\ref{boundsf})~\cite{recent}. 

Moreover,  since protons, electrons and singlet fermions are all ``charged'' under the recoil U(1)$_D$ gauge group
in our Type IIB D-foam model, unlike electromagnetism, there is a {direct} coupling of the DM singlet fermions to 
{electrically-neutral Hydrogen} atoms
Since the D-photon is massless, the cross-section for its exchange $\propto v^{-4}$, where $v$ is the relative velocity between the 
singlet fermions and the hydrogen atoms, and there is a suppression factor 
$\alpha^2 \,  \Big(n_D^{(3)} \, {R^\prime}^{-1} \Big)^2 \, (1.55 \, M_s^{-2})^{4} $ that depends on the density of D-foam~\footnote{We
recall that the size of the hydrogen atom is significantly larger than the Compton wavelength of a D-particle,
whose mass $M_s/g_s$ is in the multi-TeV range.}.

This coupling of singlet fermions to neutral hydrogen matter implies that most of the CMB constraints that are so important
in the millicharged case are avoided, as discussed in \cite{recent} (see Fig.~6 and the related discussion in that work), 
since the population of Hydrogen gas is much greater than that of the ionised matter that is relevant in the millicharged DM case.
The D-photon interactions with individual SM particles could therefore be much weaker than the interaction of the
conventional photon with millicharged particles and still provide significant baryon cooling without contradicting the 
CMB constraints. 

The quantity $\xi$ may be estimated by 
assuming that  the baryon number density $n_b (z)$ during the relevant epoch $z = {\cal O}(15)$ is associated mainly with 
hydrogen (protons), and may be 
evaluated using the Planck measurements~\cite{planck} of the current era baryon density $\Omega_b$, 
using the standard $(1+ z)^3$ redshift scaling of cosmic matter:
$n_b (z) = \Omega_b\, \rho_c \, (1 + z)^3/m_p$, where $m_p \simeq 1$~GeV,
and $\Omega_b \simeq 0.023 \, h^{-2}$,  $h=0,71$, and $\rho_c = 8.099\, h^2 \times 10^{-47}\,{\rm GeV}^4$. 
Taking a fiducial temperature for baryons $T_b \simeq 10~{\rm K} \simeq 9 \times 10^{-13}$~GeV at $z=20$~\cite{recent1,recent}, 
(\ref{logdebye2}) yields
\begin{eqnarray}\label{xidf}
\xi = {\rm ln} \Big(9 \, T_b^3 \, (4\pi \, n_b \, \alpha^3)^{-1} \Big) - 10\, {\rm ln}\,\varepsilon  ={\rm ln} \Big(7.8 \times 10^{13}\Big) \simeq  32 - 10\, {\rm ln}\,\varepsilon
\end{eqnarray}
for the Type IIB D-foam model. Assuming that the D-foam is in thermal equilibrium at the relevant epoch, 
the evolution equation for the relative velocity between baryon and DM and the cooling temperature depend on the cross-section,
as follows from the analysis in \cite{recent,cooleqs}. Constraints from CMB and Planck data can be taken into account by
following the millicharged DM analysis.
Although from this point of view the mass and the density fraction of the sterile neutrino DM are in principle
adjustable parameters, we take into account the constraints derived in~\cite{ruffini}, 
and require masses of a few tens of keV for the sterile neutrinos, which may thus play an important  r\^ole in galactic structure. 

We now examine the bounds from Planck CMB measurements~\cite{planck} on the D-photon-mediated cross-section, 
thereby constraining the D-foam parameters. 
Parameterising a generic DM-baryon cross-section as $\sigma_t = \sigma_0 v^{-4}$, 
it was demonstrated in~\cite{recent} that, for DM masses less than a few MeV, the Planck CMB constraints at redshifts $z = 17 $ can be satisfied for 
\begin{equation}\label{planck}
\sigma_0 \lesssim  1.7 \times 10^{-41}\, {\rm cm}^2 \qquad {\rm at \quad 95\% \quad C.L.}.
\end{equation}
We then find, using (\ref{sigmat2}), that
\begin{equation}\label{newepslim}
\xi^{1/2} \, \varepsilon \lesssim 10^{-10}~, \quad (m_N \ll m_b \sim 1 ~{\rm GeV})~.
\end{equation}
In the model of Type-IIB D-foam discussed in \cite{li}, the value 
\begin{equation}
n_D^{(3)} \simeq 10^{-3} M_s^{-3}, \qquad R^\prime \simeq 338 M_s^{-1}
\end{equation}  
was used for the density of D-particles, which is consistent with constraints on a possible 
foam-induced refractive index for photons at late redshifts $z \lesssim 1$, 
as well as the strong constraints on electrons coming from observations of synchrotron radiation from the Crab Nebula ~\cite{review}. 
In this case, we obtain from (\ref{neweps}) 
\begin{equation}\label{newep2}
\varepsilon \simeq  7 \times 10^{-6} \, .
\end{equation}
Using (\ref{xidf}), this implies  
\begin{equation}
\label{151}
\xi \simeq 151 \, .
\end{equation}
Comparing with (\ref{newepslim}), we see that the Planck CMB constraints~\cite{planck} on the cross-section are {not} 
satisfied for sterile neutrino DM, as the corresponding cross-section exceeds the CMB limits by several orders of magnitude. 
One would need to postulate much more dilute foam densities at redshifts $z={\mathcal O}(20)$, 
which is possible because the D-foam density is free and redshift-dependent  in general, as discussed in~\cite{dfoam}.

The constraint (\ref{151}) could be evaded if D-particles form condensates on the brane Universe, due to their stringy self-interactions,
by analogy with axion condensate formation in field~\cite{sikivie2} or string theory~\cite{axiverse}.
As discussed in \cite{vergou}, the fact that D-particles live in the higher-dimensional bulk implies that the bound states between D-particles and the brane world
do not overclose the universe, as there are contributions to the brane world vacuum energy from the bulk D-particles of mixed signs, 
depending on their distance from the brane world~\cite{westmuckett}.  As mentioned previously, coherent oscillations of a
condensate of D-particles would lead to the appearance of a vector ${\mathcal A}_\mu$ excitation
that mediates interactions between baryons and DM particles propagating in the D-foam. The formation of a D-particle condensate
would imply a coherence length that grows linearly with the scale factor of the expanding brane universe, 
strengthening D-foam-mediated effects in induced DM-baryon interactions. At the field-theory level,
this would lead to an induced ``superconductivity'' phenomenon in the interactions of the recoil-induced D-photon. 
The ``magnetic field'' of the U$_D(1)$ recoil gauge theory would be ``expelled'' from the region occupied by the D-particle condensate
\`a la Meissner, which would imply a low ``magnetic'' permeability but  high ``vacuum permittivity'' $\epsilon$ 
(i.e., a high ``dielectric constant'') for the D-foam ``vacuum'' (since their product is always $1/c^2$).
In such a case the Debye screening length (\ref{ldd}) could be significantly larger than calculated above, 
leading to larger cross-sections (see (\ref{sigmat2}), (\ref{thmin}) and (\ref{logdebye2})). 
We leave the investigation of such models to future work.

\section{Conclusions and Outlook \label{sec:concl} }

Triggered by the advent of early-Universe 21-cm astrophysics, as heralded by the first report from
the EDGES Collaboration~\cite{edges}, we have discussed in this paper the possible
interactions between baryons and singlet DM fermions such as sterile neutrinos
that could be induced by D-foam in Type-IIB string theory. As discussed in~\cite{li}, the recoil
of particles scattering off D-matter is mediated by an effective photon-like excitation, the D-photon. This
leads to a cross-section with Rutherford-like behaviour, which depends on the inverse fourth 
power of their relative velocity, implying a significant enhancement of such interactions
for slowly-moving baryons that would be most relevant during the cosmic dawn era. We have
estimated in this paper the normalization of corresponding induced cross-section between baryons and DM,
and discussed the constraints imposed by the CMB and the possible relevance for baryon cooling
at redshifts $z \sim 10$ to 20. 

The velocity dependence of D-photon exchange is the same as that of photon exchange
in the millicharged DM scenario, though here the exchanged particles are not electromagnetic photons, 
but Goldstone bosons due to Lorentz violation associated with quantum-gravitational fluctuations of the stringy 
space-time foam. The corresponding coupling depends on the density of foam, which is an
adjustable phenomenological parameter, together with the mass of the singlet fermion.
There is considerable flexibility in the magnitude of the DM singlet fermion mass, which might
itself might be generated by the interaction with the D-foam. It is suppressed hierarchically by the 
coupling $\alpha_V$ of the DM with the D-photon, whose value is poorly constrained, but may be
enhanced hierarchically by geometrical effects in a multi-brane-world scenario~\cite{MR}, as discussed
in \cite{emnsinglet}.

An important difference from the millicharged DM scenario is that both ionised and electrically-neutral 
matter particles are ``charged'' under the recoil D-matter gauge group U(1)$_D$. Hence
there is a coupling of  DM singlet fermions with the bound electrons and protons of the hydrogen atoms
that constitute most of the baryonic content of the Universe. The constraints on D-photon interactions
are therefore very different from those on millicharged DM, avoiding most of the CMB constraints~\cite{recent}. 

Detailed phenomenological studies are beyond the scope of this paper, but this exploratory study shows that 
21-cm astronomy might be relevant to the phenomenology of D-foam. We recall that this is 
characterised by a D-particle density that is redshift-dependent, in general. D-particles can
affect the refractive index of photons, which is constrained after recombination by measurements
of the arrival times of high-energy photons from intense astrophysical sources
such as Gamma Ray Bursts, Active Galactic Nuclei, along the lines suggested in previous works~\cite{emnnewuncert}.
However, these constraints do not apply to the era before recombination that is accessible to 21-cm astronomy.

We have made no claims here about the interpretation of the EDGES data~\cite{edges}, which is hotly debated~\cite{Hills}.
Rather, our aim here has been to put some ideas on the table, 
suggesting scenarios in which (part of the) dark matter/dark energy  in the Universe is of
space-time foamy origin, which might have interesting phenomenological interfaces
with future developments in 21-cm astronomy.  

\appendix  

\section{Lensing Effects of D-foam \label{sec:lens}}

\numberwithin{equation}{section}

\setcounter{equation}{0}

We comment in this Appendix on a complementary aspect of D-foam arising from its back reaction on
the space-time geometry through which baryons and other matter particles propagate.
 
As discussed in~\cite{mavdecoh}, the recoil of the massive D-particle defects  during their encounter with a propagating particle 
causes the space-time `seen' by the baryon at large scales to appear topologically non-trivial, deviating
from Minkowski by a small solid-angular surplus: 
\begin{equation}\label{deficit}
ds^2 = dt^2 - dr^2 - b^2 \, r^2 ({\rm sin}^2\theta \, d\phi^2 + d\theta^2), \quad b^2 = \frac{1}{1 - \Delta^2}, \quad \Delta^2 \ll 1 \, , 
\end{equation}
in standard spherical polar space-time coordinates, where 
\begin{equation}\label{deltarec}
\Delta^2  = < |\vec u|^2 >_{\rm recoil} = g_s^2\, \sigma^2 \frac{|\overline{\vec p}|^2}{M_s^2}
\end{equation}
is the variance of the recoil velocity of the D-particle of mass $M_s/g_s$, the average
$< \dots >$ is taken over the statistical ensemble of the D-particle defect populations,
and $\overline{\vec p}$ denotes an average particle momentum.

Assuming a momentum transfer during an individual scattering $\Delta p_i = r_i p_i$, 
$r_i < 1$, $i=1,2,3$ (no sum over $i$), then $\sigma^2$ denotes 
the stochastic variance for $r_i$, of the form  $<r_i r_j>=\sigma^2 \delta_{ij}$, $<r_i>=0$.
In the case of non-relativistic thermalised baryons
 in a heat bath of temperature $T$, their average (thermal) momentum is 
 $|\vec p|^2 \sim 3 m_b \,T$, and (\ref{deltarec}) leads to
 \begin{equation}\label{deltarec2}
 \Delta^2  = < |\vec u|^2 >_{\rm recoil}\,  \sim \, \frac{3 \sigma^2 m_b \, T}{M_s^2} g_s^2~, 
 \end{equation}
where $\sigma^2 < 1$ a phenomenological 
stochastic fluctuation parameter of the D-foam, that sets the scale for the `fuzziness'  of space-time. 

If one assumes that the ensemble of D-particles on our brane Universe are themselves thermalised at a temperature $T$, 
at least at late eras of the Universe~\cite{sakel}, then one can define the following stochastic parameter related
to their thermal velocity:
\begin{equation}\label{dfoamtherm}
\delta^2_T = < {|\vec u ^{{\rm(D)}}}|^2>_T \, , 
\end{equation}
where $\vec u^{{\rm (D)}}$ is the thermal velocity vector of a D-particle with mass $M_s/g_s$. 
The thermal velocity  variance can be then estimated using the thermodynamic relation  
$<\frac{M_s}{g_s} {|\vec u^{(D)}|}^2 >_T = 3 T$, yielding 
\begin{equation}\label{delta}
\delta^2_T = g_s  \frac{3\, T}{M_s}.
\end{equation}
We note that this is much larger than the statistical fluctuations due to recoil momentum transfer 
under the same thermal conditions, shown in (\ref{deltarec2}),
since $m_b \ll M_s/g_s$. 
For temperatures $T \sim 10 ~{\rm K}$ typical of the cosmic  dawn era and $M_s/g_s > 10~ {\rm TeV}$ 
(as implied by current collider limits from searches for extra dimensions) we have
\begin{equation}\label{deltaestimate}
\delta^2_T < 2.6 \times 10^{-16}, \quad \Delta^2 <  2.6 \times 10^{-20}\, \sigma^2, \quad \sigma^2 < 1 \, .
\end{equation}
Thus there are at least four orders of magnitude difference between these two parameters of the D-foam~\footnote{The 
reader should notice that for relativistic particles, such as photons, for which the average thermal momentum 
scales as $|\vec p| \sim T$, there is a much stronger suppression of the parameter $\Delta^2$ 
compared to $\delta_T^2$, by at least sixteen orders of magnitude in the conditions considered here.
Hence the slow-moving baryons provide much more sensitive probes of the foam parameter $\Delta^2$, and
the topologically non-trivial structure of `fuzzy' space-time (\ref{deficit}).}.

As discussed in \cite{mazur,mavpap}, the propagation of particles in topologically non-trivial space-times of the form (\ref{deficit})
results in non-trivial scattering, whereby the defect `lenses' the particles, in the sense of inducing singular 
scattering amplitudes for scattering angles $\theta = \pi (1 - b^{-1})$. Such cross-sections are 
not of Coulombic Type, in the sense of scaling as $|\vec p|^{-2}$ rather than $|\vec p|^{-4}$, where $|\vec p|$ 
is the magnitude of the spatial momentum 
of the particle in the frame where the defect is at rest,
in contrast to the standard millicharged DM interactions that scale like $v^{-4}$.
As was carefully demonstrated in \cite{mavpap}, these cross-sections vanish in the no-defect limit
so, for sufficiently small $\Delta^2 \ll 1$, the corresponding differential cross-sections are suppressed
compared, for instance, with the cross-sections discussed in the main text of this paper. Nonetheless, 
such an effect should in general be taken into account, especially in Type-IA models of D-foam, 
where there is no direct interaction of baryons with the D-particles. In such models the baryons
simply feel the effects associated with the non-trivial topology of the `fuzzy' space time (\ref{deficit}) 
via the induced cross-sections of \cite{mazur,mavpap}. In addition, in such space-times, the 
separation of the energy levels of atoms such as hydrogen, of relevance to us here, 
will be modified compared to the non-defect limit~\cite{hydro}. Thus, constraints on the strength of
$\Delta^2$ could in principle be imposed in the framework of the 21-cm astronomy.
However, the parameter $\sigma$ is expected to be strongly suppressed, so indirect constraints 
on other parameters of the D-foam, as discussed in the main text, are expected to be much stronger. 

We illustrate the above ideas by considering the scattering of a fermion in a space-time (\ref{deficit}), 
as studied by Ren in \cite{mazur}. The scattering amplitude induced by a defect is given by
\begin{equation}\label{spin}
f(\theta) = \frac{i}{|\vec p|} \delta(1 - {\rm cos}\theta) + \frac{{\rm sin}(\pi \tilde \alpha) \, (1 - {\rm cos}\theta)^{1/2}}{2\sqrt{2}\, |\vec p|\, ({\rm cos}(\pi \tilde \alpha) - {\rm cos}\theta)^{3/2}}, \quad \tilde \alpha \equiv 1 - b^{-1}~,
\end{equation}
and the corresponding differential cross-section is
\begin{equation}\label{diffspin}
\frac{d\sigma}{d\Omega} = |f(\theta)|^2 =   
\frac{{\rm sin}^2(\pi \tilde \alpha) \, (1 - {\rm cos}\theta)}{8\, |\vec p|^2\, ({\rm cos}(\pi \tilde \alpha) - {\rm cos}\theta)^{3}} + \dots  \, .
\end{equation}
where the $\dots$ represent singular terms arising from 
the $\delta$-function term in the scattering amplitude (\ref{spin}), which need proper regularization in the $\theta=0$ (forward scattering) region, 
as discussed in \cite{mavpap}. This is essential for yielding the correct vanishing result for the scattering amplitude
in the no-defect limit, where $f(\theta)\,{|_{\rm no-defect}}$=0, and hence the suppression of the cross-sections for vanishingly-small deficits.
Notice also the singularity when $\theta=\pi \tilde \alpha$ of the second term in (\ref{diffspin}), which 
is independent of the spin of the particle~\cite{mazur}, and leads to the lensing effect on the particle. 

The integration over the scattering angle $\theta$ encountered in the calculation of the total cross-section
exhibits singular behaviours in (\ref{diffspin}) at the values of the scattering angle 
$\theta=\{0,\pi\tilde \alpha \}$ due to the first and second terms of (\ref{diffspin}), respectively.  
For the small $\pi \tilde \alpha \ll 1$ that characterise our D-foam, 
one may impose a cut-off $\theta_{\rm min}$ for the $\theta$ angle to prevent it from approaching zero:
$\theta_{\rm min} > \pi\tilde \alpha \ne 0$, which automatically regularises the $\delta$-function term in (\ref{diffspin}). 
However, this still leaves the other 
singular limit $\theta \to \pi\tilde \alpha$. In this case, we may regularise the denominator of the second term 
on the right-hand side of (\ref{diffspin}) by means of an angular resolution cut-off $\mathcal \Delta$, 
independent of $\pi \tilde \alpha$, such that
\begin{equation}
{\rm cos}(\pi\, \tilde \alpha) - {\rm cos}\theta \,  \stackrel{\theta \to \pi\tilde \alpha}{\simeq} \, {\mathcal D}^2  \ll 1 \, .
\end{equation}
In our foam case, we can approximate $\pi\tilde \alpha  \simeq \frac{\pi}{2}\Delta^2 \ll 1$, 
so that ${\rm cos}(\pi\tilde \alpha) \simeq 1 - \frac{(\pi\tilde \alpha)^2}{2}$. As we demonstrate below, 
this is a self-consistent approach that leads to finite and well-behaved cross-sections, 
which vanish in the no-defect limit, as discussed in \cite{mavpap}.

The angular resolution $\mathcal D$ is assumed to be provided by the thermal motion of the D-foam, 
in the sense of the uncertainty in the scattering angle induced by the thermal fluctuations of the scatterer,
which are considered much stronger than any inherent quantum-gravity effect. Hence, for our purposes, we can identify 
\begin{equation}\label{defcalD}
{\mathcal D}^2= \delta_T^2 = g_s  \frac{3\, T}{M_s} \, ,
\end{equation}
with $\delta_T^2$ given in (\ref{delta}),  with
\begin{equation}\label{ineq}
1 \gg {\mathcal D} \gg  \pi \, \tilde \alpha =\frac{\pi}{2}\Delta^2
\end{equation}
in view of (\ref{deltaestimate}).

For the calculation of the corresponding transport cross-section 
(\ref{sigmat1}) of relevance to us here, it suffices to ignore the $\delta$-function term in (\ref{diffspin}),
which can be easily justified by assuming a 
minimum cut-off for the angular integration variable $\theta > \pi \tilde \alpha$). 
Concentrating on the second term, we may make a formal change of integration variable 
to $y={\rm cos}(\pi\tilde \alpha) - {\rm cos} \theta \simeq  1 - {\rm cos}\theta - \frac{(\pi\tilde \alpha)^2}{2}$, 
and write (by taking into account that $\theta \in (\pi \tilde \alpha, \pi]$)
 \begin{equation} 
  \sigma_t = \int_{\frac{(\pi\tilde \alpha)^2}{2}}^{-1} d({\rm cos}\theta)\, (1 - {\rm cos}\theta) \frac{d \sigma(\theta)}{d({\rm cos}  \theta)} \, \simeq \, \frac{(\pi \alpha)^2}{8 |\vec p|^2} \, \int_0^2 dy \frac{\Big(y + \frac{(\pi\tilde \alpha)^2}{2}\Big)^2}{y^3} \, ,
  \end{equation} 
to leading order in $\pi\tilde \alpha$. As already mentioned, we observe that this leads to a formal singularity 
for the value $\theta =\pi\tilde \alpha$ due to the `lensing' effect, which requires additional regularisation. 
According to our previous discussion, we now regularise the lower $y$-integration bound by replacing 
$0$ with ${\mathcal D}^2$ (\ref{defcalD}). 
We then easily obtain for the transport cross-section (\ref{sigmat1}):
\begin{eqnarray}\label{sigmatfuzzy}
&& \sigma_t = \int d({\rm cos}\theta)\, (1 - {\rm cos}\theta) \frac{d \sigma(\theta)}{d({\rm cos}  \theta)} \, \stackrel{\Delta^2 \ll 1}{\simeq} \, 
\frac{(\pi \alpha)^2}{8 |\vec p|^2} \Big[ {\rm ln}\Big(\frac{2}{{\mathcal D}^2}\Big) +  (\pi \alpha)^2 \Big(\frac{1}{{\mathcal D}^2} - \frac{1}{2}\Big) +
\frac{(\pi \alpha)^4}{8} \, \Big(\frac{1}{{\mathcal D}^4} - \frac{1}{4}\Big)\Big] 
\nonumber \\
&& \simeq \frac{\pi^2 \Delta^4}{32 \,m_b^2} \, {\rm ln}\Big(7.6 \times 10^{15} \Big) \, \frac{1}{v^2} + \dots \simeq 1.07  \, \frac{\pi^2 \Delta^4}{m_b^2} \, \frac{1}{v^2} + \dots \, ,
\end{eqnarray}
where $\dots$ denote subleading terms, $m_b $ is a generic baryon mass and
 $v$ is a relative velocity of the baryon with respect to the D-partlcle scatterer.
Given that $\Delta^2$ is sufficiently small (\emph{c.f.} (\ref{deltarec2}), (\ref{deltaestimate})), 
we observe that this is suppressed compared to (\ref{sigmat}) as we mentioned previously, despite the $v^{-2}$ behaviour. 

We note for completeness that, with our regularisation, the total cross-section 
$\sigma_{\rm total} = \int d\Omega |f(\theta)|^2 $ is
\begin{equation}
\sigma_{\rm tot} 
 \simeq \frac{(\pi \alpha)^2}{8 |\vec p|^2} \, \int_{0 \to \mathcal D^2}^2 dy \frac{y + \frac{(\pi\tilde \alpha)^2}{2}}{y^3}
\simeq 
\frac{(\pi \tilde \alpha)^2}{8 |\vec p|^2} \Big[ 
\frac{1}{{\mathcal D}^2} - \frac{1}{2} + \frac{(\pi \tilde \alpha)^2}{4} \, \Big(\frac{1}{{\mathcal D}^4} - \frac{1}{4}\Big)\Big] \, , 
\end{equation}
which is finite like the transport cross-section and, as with $\sigma_t$, also
vanishes in the no-defect limit $\tilde \alpha \to 0$, providing a self-consistency check of the approach~\cite{mavpap}.

In Type-I D-foam models, where the baryons do not interact directly with the foam, 
the effect (\ref{sigmatfuzzy}) is the dominant effect of the foam on baryonic matter. In such a case, 
one obtains a transport cross-section of the form $\sigma_0 v^{-2}$, which may contribute to baryon cooling. 
From CMB limits, following the analysis of the second paper in~\cite{recent}, we know that with such a
velocity dependence of the transport cross-section between baryons and dark matter,  one has
 $\sigma_0  \lesssim 2.3 \times 10^{-33}~{\rm cm}^2$ at 95\% C.L., implying on account of (\ref{sigmatfuzzy}) that
 \begin{equation}\label{cmblimi}
  \Delta^2 \,   \, < \,   7.5 \times 10^{-4} ~, \quad {\rm at ~95\%~C.L.} \, .
 \end{equation}
 As already mentioned, this constraint can be combined with the effect on hydrogen hyperfine splitting of the 
 topologically non-trivial space-time (\ref{deficit})
following the analysis in \cite{hydro}, which will in general modify the features of the 21-cm emission spectrum. 
The limit (\ref{cmblimi}) is much weaker than the theoretically expected bound (\ref{deltaestimate}). 
 On the other hand, had we used the constraint on $\sigma_0$ imposed by CMB considerations~\cite{recent} 
 on Coulombic cross-sections $\sigma_0 v^{-4}$ between DM and baryons, $\sigma_0 < 1.7 \times 10^{-41}~{\rm cm}^{-2}$, 
 we would have derived an experimental bound on $\Delta^2$ four orders of magnitude stronger than (\ref{cmblimi}), 
 though still much weaker than the theoretical value (\ref{deltaestimate}). It is for this reason that the consideration
in the main text using the foam as a mediator between baryons and particle DM, lead to much stronger 
bounds on the relevant foam parameters, such as its density at a given redshift.

However, there are several steps that they need to be taken into account before the above ideas 
acquire a concrete shape and lead to a detailed phenomenology. These include a detailed study of the 
effects of D-foam with the above parameters on the CMB spectrum and the associated polarisations, 
given that the photons do interact with it, the effect of the drag on baryons on Baryon Acoustic Oscillations, 
and additional astrophysical sources that may heat the baryons, counteracting their  cooling and hence introducing 
uncertainties in the interpretation of the signal reported by EDGES~\cite{recent}, etc.

\section*{Acknowledgements}

The work of JE and NEM is supported in part by the
U.K. Science and Technology Facilities Council (STFC) via the Grant ST/L000258/1,  
and that of JE is also supported in part by the Estonian Research Council via a Mobilitas Pluss grant.
NEM also acknowledges the current hospitality of IFIC Valencia through a Scientific Associateship (\emph{Doctor  Vinculado}).
The work of DVN is supported in part by DOE grant DE-FG02-13ER42020 and
in part by the Alexander S. Onassis Public Benefit Foundation.

\end{document}